\documentclass{easychair}

\usepackage{doc}

\usepackage{amssymb}
\usepackage{hyperref} 

\usepackage{paralist}

\title{IPPO: A Privacy-Aware Architecture for Decentralized Data-sharing}

\author{
    Maurizio Aiello\inst{1}
\and
   Enrico Cambiaso\inst{1}
\and
    Roberto Canonico\inst{2}
\and
    Leonardo Maccari\inst{3}
\and
    Marco Mellia\inst{4}
\and
    Antonio Pescap\`e\inst{2}
\and
    Ivan Vaccari\inst{1}
}

\institute{
  National Research Council (CNR-IEIIT), Italy
  \email{{name.surname}@ieiit.cnr.it}
\and
  University of Naples Federico II, Italy
  \email{roberto.canonico@unina.it,pescape@unina.it}
\and
  Universit\'a C\'a Foscari di Venezia, Italy
  \email{leonardo.maccari@unive.it}
\and
  Politecnico di Torino, Italy
  \email{marco.mellia@polito.it}
}

\authorrunning{Aiello et al.}

\titlerunning{IPPO}

\begin{document}

\maketitle

\begin{abstract}
Online trackers personalize ads campaigns, exponentially increasing their efficacy compared to traditional channels.
The downside of this is that thousands of mostly unknown systems own our profiles and violate our privacy without our awareness.
IPPO turns the table and re-empower users of their data, through anonymised data publishing via a Blockchain-based Decentralized Data Marketplace.
We also propose a service based on machine learning and big data analytics to automatically identify web trackers and build Privacy Labels (PLs), based on the nutrition labels concept.
This paper describes the motivation, the vision, the architecture and the research challenges related to IPPO.
\end{abstract}


\section{Introduction}

Online advertising is characterized by a constant growth and it generated \$49B in revenues in 2014, considering US alone \cite{korula2016optimizing}.
Advertisements (ads for short) support important Internet services such as search engines, social media and user generated content sites.
The ability to target individuals with specialized advertisements makes online ads success.
For this reason, different tracking techniques are commonly adopted to profile each user and identify
their interests based on their browsing history. As a result, the web is full of
hidden tracking services that extract data from our online activity \cite{Falahrastegar2014Rise}.
Although the recent introduction in Europe of the General Data Protection Regulation \cite{de2016new} is a strong attempt to protect users' privacy,
we steadily leak very personal information, while covert web companies build a
profile of each of us, and sell information they collect \cite{Metwalley2015Online}.
The recent Facebook-Cambridge Analytica \cite{cadwalladr2017great} scandal is just the top of the iceberg.
The creepiest part which still sits under the sea consists of obscure tracking
services, run by thousands of mostly never heard companies, which keep
collecting data when we visit any website, from any device, from any place.
They act as data brokers, monetising our profiles.
Under another perspective, we observe that today web browsing is incomparably
more complicate than only a few years ago. Web services are utterly complex,
browsing platforms are extremely variegate and the attention span of users is
very low, so that a web service to be successful must be intuitive and of
immediate use. Service providers need data to design their services and to
tailor them to the specific requirements of a single user. Therefore, data
gathering is not only connected with privacy-invasive ads campaigns, it is
needed to offer competitive web services on the Internet ecosystem in general.

This leads to two fundamental questions this paper address: \emph{is it
possible to offer an economically sustainable alternative to the model made of
(mostly) free services and privacy-infringing  advertising?}
And, \emph{if we find an alternative to data brokers, how will on-line services
collect enough data to tailor their on-line services?}

To give an answer we can not operate only at the technical layer, we need to
conceive new ecosystems that enable a fair exchange of data from the producers
(the Internet users) to the consumers (the service providers that need to
optimize their services) preserving both privacy and service availability.
IPPO is an architecture that encompasses all the necessary components to propose
an alternative: it monitors the behaviour of web services to deter the most
intrusive forms of tracking, but it also enables users to create eventually anonymised
data that they can willingly share with service providers. The service
providers in turn can acquire data from the users (and with their explicit consent) to understand the user habits and
improve their services, using a transparent and privacy-aware platform.

The rest of the paper is organized as follows: \ref{sec:motivations} goes deep
into the motivations of IPPO, \ref{sec:architecture} explains the architecture
we envision and \ref{sec:components} details all its components, \ref{sec:BDG}
focuses on the Big Data Grinder, an example application to run on IPPO. Later on
\ref{sec:impact} discusses the impact we expect IPPO to have, and
\ref{sec:agenda} introduces the research challenges we need to address to
complete IPPO. Finally, \ref{sec:conclusions} draws conclusions.

\section{Motivations}
\label{sec:motivations}

We are nowadays used to receive targeted advertisements. After googling for a pair
of sunglasses, automatically, sunglasses advertisements will pop-up in our
navigation on websites we do not directly relate to Google, and e-shops will
suggest us new pairs of sunglasses we did not consider before.
The more tech-savvy people simply accept that in exchange of free services, we
give away personal information, while the layman simply ignores to be part of
this system.
The only possibility to rebalance this flow of information today completely
dominated by Internet Corporations is to give back to the users the ownership and
control over their data, along with the choice and the freedom to take informed decisions,
and tools to enforce policies at their will.
The first step is to make it evident for them that data have a value, and that
free on-line services are free, as long as we do not consider the value of the
data that we give away.
IPPO is grounded on the principle that people should be compensated for sharing
their data. If people understand that data have an economic and tangible value,
they will hardly give data away, and they will be actively engaged in protecting
their data. Monetising data will finally add transparency in the
web tracking ecosystem.  

On the other hand, web services need user data to design, run and improve the service they offer.
Top service providers (such as Google, Facebook, Amazon to name a few) nowadays are in a position to directly collect users' data. Smaller companies can instead buy this data from the data brokers that offer them user profiles. 
Data are used for a variety of reasons, to offer personalized content, to improve user experience, or to give geolocalized information, but also to train anti-spam algorithms or intrusion detection systems.
IPPO wants to provide eventually anonymised data sources for data consumers, i.e., those who need data to improve their systems, bypassing the data brokers, and under the full control of the data producers (i.e., the end users). Data-sets will be produced and shared by the end user, e.g., using a browser plugin, and publicly sold in a distributed marketplace. 
The data consumers will access this marketplace and buy the data they need for
their goals. 
The user will be compensated for sharing data, and targeted services may be
acquired. No intermediary is needed, and no aggressive user profiling is
required. Everything will be consolidated using smart contracts.

Targeted advertisement requires user profiling, and thus our architecture lets end user to control whether to offer eventually anonymised data. This may disfavour the targeted
advertising industry. Indeed,  we strongly believe that user profiling is an approach that implies too many negative externalities to be socially sustainable on the long run, and we propose a viable alternative with a healthier approach. 

\section{IPPO Overall Architecture}
\label{sec:architecture}

\begin{figure*}
\begin{center}
\includegraphics[width=0.6\linewidth]{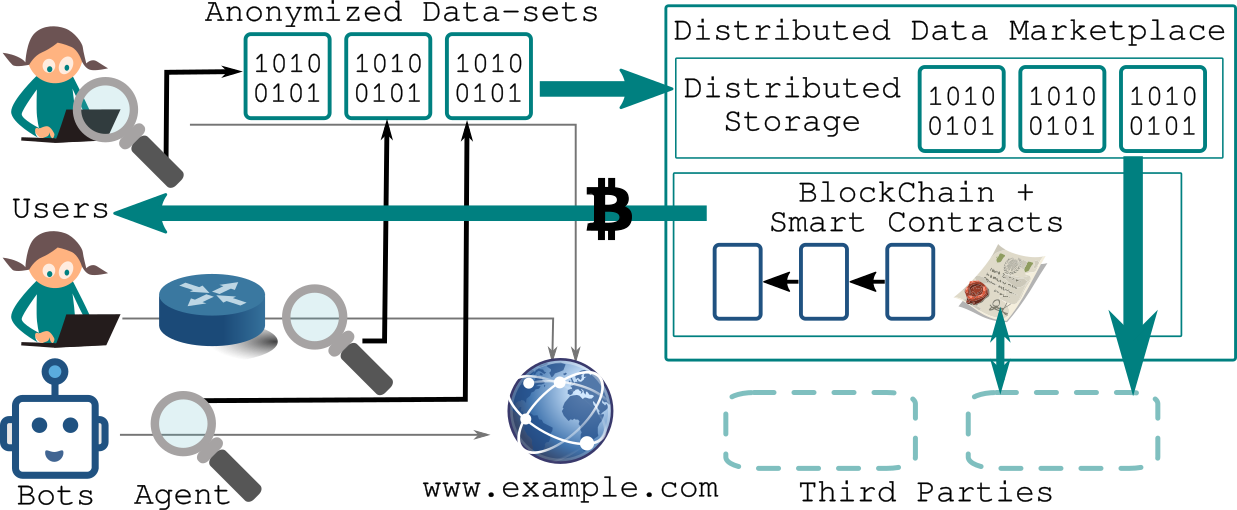}
\caption{The overall architecture of IPPO.}
\label{fig:architecture}
\end{center}
\end{figure*}

The IPPO architecture is depicted in \ref{fig:architecture} and is made of three main components: 
\begin{inparaenum}[(i)]
\item IPPO Agents perform data collection, anonymisation and sharing. Agents are
envisioned as plug-ins that directly run in the users devices, and/or middleboxes placed as, e.g., virtual network functions in corporate networks.
\item a Distributed Data Marketplace (DDM) is an open, decentralized storage
platform in which users publish anonymised navigation traces, combined to a
Blockchain to empower a real data marketplace. The DDM offers any third party
the ability to ``buy\&sell'' data using digital contracts.
\item third Parties can acquire data-sets to perform data-analysis on the data-sets.
\end{inparaenum}

As an example of third-party application we propose the Big Data Grinder (BDG),
a system that runs scalable machine learning based algorithms to automatically identify
traffic from/to trackers (later on referenced to as ``tracking traffic'').
The BDG performs automatic auditing of web
services, and identifies privacy threats or involuntary leakage of personal
information. 
The BDG produces feedback to the users in two forms:
\begin{itemize}
\item Privacy Labels and Privacy Factor: graphical representation to communicate
to non-experts the privacy cost of online services. Privacy Labels translate the
results of the auditing algorithms into simple and tangible metrics (better explained in \ref{sec:BDG}) to indicate both which information a website collects,
and the cost of accessing it.
\item Filtering Rule-Sets: the BDG can produce also rule-sets that can be enforced
directly on the clients machines. Agents could even support specific plug-ins
from trusted third parties to perform specific actions.
\end{itemize}
The next sections will describe in details each component, and the related
research challenges.


\section{IPPO Components}
\label{sec:components}

\subsection{IPPO Agents}
Data is initially collected by the IPPO Agents, of three kinds:
\begin{inparaenum}[(i)]
\item browser plug-ins or mobile apps, which the user freely installs. They access unencrypted
data even in HTTPS sessions, allowing total visibility on web traffic;
Data is collected in anonymised form, thanks to anonymisation techniques like
differential privacy that make it possible to share such raw data, while
guaranteeing end-user privacy. Data-sets are encrypted, annotated with
meaningful automatically generated meta-data, stored in the DDM, and announced
on the Blockchain (described later on).
\item A companion instrument to user agents are automatic bots, which scrape the
web at scale. Bots can form a basis for a wide data-collection, and can also
reproduce user browsing, so that once a browser agent has realized a traffic
trace, the agent can trigger one of the available active probes (headless
browsers) to partly repeat the browsing session. In this case, new data-sets
related to the initial one can be published in the DDM.
\item Agents can also be installed on middleboxes and from privileged positions (like corporate internal proxies) they generate large, aggregated traffic data-sets.
\end{inparaenum}

\subsection{The Distributed Data Marketplace}
\begin{figure}
\begin{center}
\includegraphics[width=0.7\linewidth]{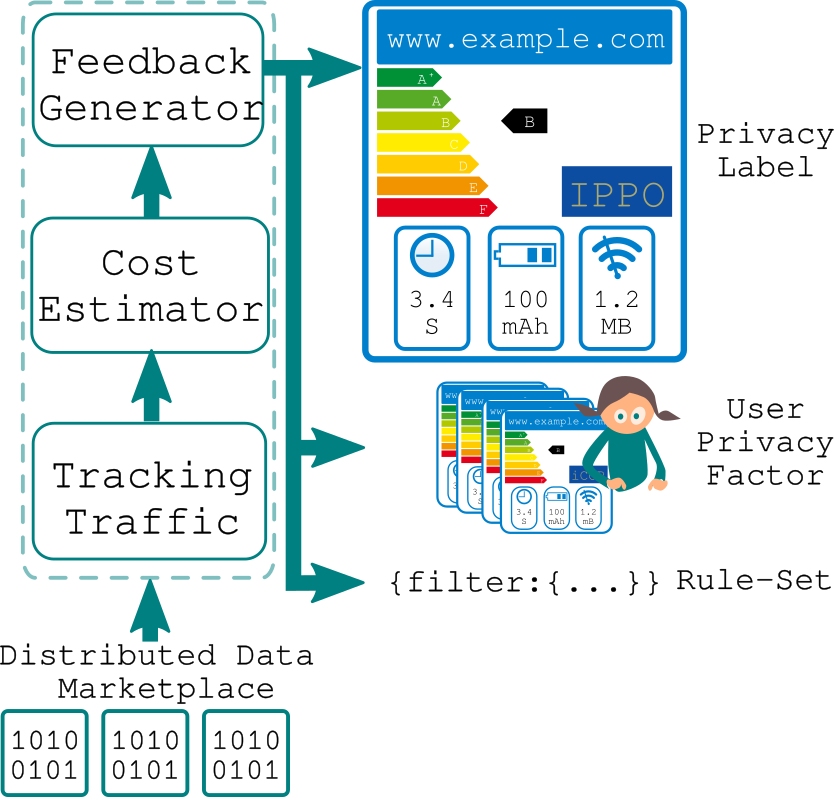}
\caption{The architecture of the BDG.}
\label{fig:bdg}
\end{center}
\end{figure}

The concept of DDM rebalances the current disparity of power on the Internet:
instead of a handful of big players with their walled gardens, an open
marketplace where users can make their data (literally) count. 
The DDM is made of two components:
\begin{itemize}
\item IPPO Distributed Storage: it provides a decentralized storage for the
encrypted data-sets. IPFS could be the natural choice for this component, as it
enables a truly decentralized and cryptographically robust file system imagined
to scale to millions of nodes. Users will store their anonymised data-sets in
the storage, encrypted with a unique key, which will be revealed only to the
buyer of the data. After the acquisition the data-set can be removed from the
storage, or left in the storage for future transactions.
\item Blockchain with Smart Contracts support: when a user generates a new
data-set from a browsing session they also add references and
detailed meta-data, automatically extracted from the traffic session, in the DDM
Blockchain. Smart contracts represent the gateway to access
the data-sets. Third parties negotiate with the smart contract the access to the
data-sets in exchange of cryptocurrency for the owner.
\end{itemize}
The DDM must be anonymous so that third parties can not correlate different
browsing sessions of the same user (unless she allows it). Techniques like
non-interactive zero-knowledge proofs will be the key enablers to realize the
DDM.

\subsection{Third Parties}
One key observation is that many market sectors for on-line services can be
classified as oligopolies. Search engines and security appliances are only two
examples in which a handful of players take the vast majority of the market. 
Big players enjoy a dominating position also because they have access to
enormous databases that they use to train and optimize their services. 
With IPPO we envision that interested third parties will use the DDM to
access to data-sets that are on the market. 
We believe that the DDM will make it possible also for newcomers and start-ups
to access enough data to start their business and re-balance market sectors that
are pathologically unfair. 
Third parties interested in these
data-traces may be of many kinds. For instance, most of the security appliances
on the market today include algorithms for identifying malicious traffic, and
need to be trained on existing data-traces. Similarly, software providers that
specialize in user experience need to have detailed traces of user actions when
using specific popular services. 

One specific third party we detail in the rest of the paper is the BDG, a
platform that can analyse data-traces and give back to the users understandable
feedback on the privacy level of their browsing habits, plus, rule-sets for
filtering out invasive trackers.
We describe the BDG in the next section.

\section{The Big Data Grinder}
\label{sec:BDG}


One of the most troubling issue with on-line trackers is the complete lack of
transparency and the sense of disempowerment that users feel when they realize the
amount of information they unwillingly give away while browsing. This is even
worse in the corporate environment, in which employees daily disclose protected
information that is collected by data brokers. A data broker can, for instance,
inform company X that the employees of its competitor company Y daily search for
job offers, and that a number of them is leaving the company.

The BDG tries to produce an informed feedback to users, made of a privacy score
for the services they use and also for their browsing habits. 
We propose to revive a concept that was introduced several years ago: the idea of Privacy Labels (PLs). Originally, they were conceived for illustrating in a simple way the service terms -- the EULA --  to users \cite{Kelley2009Nutrition}. Here we instead propose to use them to inform users about the service they contact.
PLs are similar to nutrition or energy labels, but for privacy.  They contain
indications about the trustfulness of web services and can be used as a public
``privacy fingerprint'' of an online service. How to form and fill information in
Privacy Labels is however far from being easy.  In our vision, the BGD elaborates the data-sets in order to extract from the
user traces the tracking traffic, and produces simple metrics that the common
people can understand, e.g., number of contacts with trackers, cost in data
volume, battery consumption, etc. 
In \ref{fig:bdg} the BDG is broken down in all its components; a first
module applies machine learning algorithms to automatically identify tracking traffic and a
second module  analyses this traffic and estimates the impact of such
traffic in terms of understandable metrics. These metrics take into
consideration the time that was spent to load unwanted contents, and thus the
usability impairment generated by tracking traffic, the amount of Bytes of
unwanted traffic that the user had to download, and the consequent depletion of
battery, especially useful on mobile devices. In general, these are only some of
the metrics that can be used to characterize the impact of tracking traffic in a
tangible way. In different contexts (for instance, the corporate context) some
other metrics can be of interest and could be included in the privacy labels.

In this model, the BDG acquires data-sets from the DDM and publishes PLs for
users that want to know what is the ``privacy fingerprint'' of a specific
service. The final goal of this strategy is to increase (or create) awareness in
the users about the services they are using, and helping the interested users in
choosing the one (among the available ones) that is more respectful of their
privacy.

The BDG can offer also personalized services, analysing traffic from the
browsing sessions that a customer makes available. In this case it is the
customer that makes available its data traces and potentially pays for a
personalized service. 
We envision two of these services, first, the BDG can produce per-user labels,
showing how trackers impacted user navigation. When aggregated on a per-user
base, they become the Privacy Factor, which measures the exposure of the user to
web tracking and the incurred cost.
Second, the BDG could produce rule-sets that can be given back to the users to
enforce filtering of tracking traffic. These rule-sets can be used by 
security services like firewalls, antispam filters,  DDoS mitigation services
etc. This architecture is depicted in \ref{fig:bdg} which summarises the
architecture of the BDG in all its components. 

\section{Impact and Outlook}
\label{sec:impact}

As we already mentioned we are perfectly aware that a large part of the Internet
ecosystem today is financed by the ads industry. Weather this is an economically
sustainable model, or it is just an armed race to the best and most precisely
targeted advertisement campaign, it is under discussion \cite{Blake2015Consumer}.
What is instead becoming clear is that basing the Internet economy on profiling
people is becoming socially unsustainable \cite{Schneier2015Data}.
Our ultimate goal is to propose a model enabling a different cash flow (or
\emph{cryptocash flow} for whatever difference it may make). Money, instead of going
from advertisers to web-services will flow from data consumers to the users and
then back to tailored services. Those that need the data will acquire them
directly at the source, the Internet users. Users will then spend their
resources in services. The more data they share, the more currency they own, the
more personalized services they will be able to acquire.

Of course this does not preclude the possibility of using ads, if a user wants
to receive targeted advertisement, they will be able to share data about
their shopping preferences or search results, and let third party companies to
produce tailored feedback. The main difference here, is that users will be aware
that data have an economic value, and they will decide to trade data for money,
and not passively accept to be profiled without even knowing this happens.

In this design, the BDG perfectly fits as a companion component that unveils in
an understandable way the trade-offs of the current system.  The user
needs knowledge of her data flow in order to defend her right to privacy.
Due to current web technologies, gathering such knowledge is
beyond the layman possibilities and often precluded also to the expert: the aim
of the BDG is to reduce such obscurity. Moreover, the current lack of clear
exposition of key privacy-related characteristics of the services gives the user
a blurred picture of the offer, with little or no apparent choice in matter of
privacy, effectively downplaying its importance. The DBG will contribute to a
raise awareness on users' right to privacy protection, and produce usable and
tailored filters they can actually enforce.

\section{Going beyond the SoA: a Research Agenda}
\label{sec:agenda}
While the IPPO goals and architecture are very ambitious, it is based on a solid background
of existent research, which must be improved and organized in a single system.
This section reviews the relevant state of the art and outlines the challenges
that still need to be solved to arrive to a real implementation of IPPO.

\subsection{Data Collection}

IPPO agents need to collect data in real time and anonymise it before publishing
it on the DDM.  One way of doing this is using browser plug-ins or Apps that can
access the traffic in real time. While this has been realized in the past,
\cite{Metwalley2015Crowdsurf} making data collection scale is not trivial. 
Anonymising data is even more complex, as anonymisation should be ``future
proof'' and thus, the recent advancements in differential privacy are the basis
for the realization of the IPPO agents \cite{Ebadi2015Differential}.
\subsection{Realizing the DDM}
The DDM is a complex system comprising a distributed data sharing platform and a
blockchain with support for smart contracts.  While there are working models for
both these components, they also bring a relevant amount of complexity.
Coupling blockchains with a distributed data storage has been recently proposed
\cite{Chen2017Improved,Ali2016Blockstack}, at it is at
the fundamentals of the StorJ solution, proposing an Open Source solution for
to decentralize cloud data storage, as
well as of the Filecoin cryptocurrency.  Although such approaches are particularly interesting,
they are prone to scalability issues, and the unique combination of requirements
of the DDM (in terms of privacy for the users, support for economic
transactions, robustness, etc.) makes its design a real research challenge.
Users will need to store their data-sets in a fully anonymous blockchain (and
not only pseudonymous as in most of the mainstream blockchains used in the real
world). Proposals based on zero-knowledge proofs, as in the zk-SNARK protocol
recently made their way to working systems (as in the Zcash system
\cite{Peck2016Blockchain}). But yet mixing full anonymity, distribution and
protection from data tampering is an open issue.  
\subsection{The BDG} 

Identifying tracking traffic, and data trackers is still a completely open
challenge.  Most of current solutions rely on manually pre-computed list of trackers, and
very little research has been done in the automatic identification of trackers.
A promising approach is to apply supervised machine learning to the analysis of
DOM structure \cite{Bau2013Promising}, or HTTP requests
\cite{Li2015Trackadvisor}, or fingerprinting \cite{Cao2017Browser}.  Similarly,
\cite{Ikram2017Towards} manually examines 2,600+ javascript tracking tools, and
designs a supervised classifier to automatically flag abuse. 
Given the unknown nature of tracking services, the real challenge is to  
primarily rely on unsupervised machine learning algorithms, such as clustering
and text mining, to process traces and highlight tracking traffic. Not only
traffic itself can be analysed, but also web page code (javascript in particular) 
to look for
possibly malicious code inserted to fingerprint users and exfiltrate data. 

Another open challenge is how to define methodologies and models to measure and
extract the cost of tracking traffic from labeled traces. One way to do this is
comparing the access to the same service with and without third-party services,
which is a strategy that can extract i) the extra data usage cost due to
tracking, ads or analytics services, ii) the QoE impairments, e.g., the extra
cost to load/render a web page, iii) Energy consumed by mobile terminals.
Re-browsing services with IPPO agents is a viable solution for those
services that can be browsed without private credentials, but we need to define
machine learning models to generalise and predict the cost on other web
services. 

\section{Conclusions}
\label{sec:conclusions}

This paper describes IPPO, an architecture for giving back to the users the
control on their data. We believe that the only way to rebalance the current
disparity of power on the Internet is to gradually abandon the ads-based revenue
model that dominates the market of on-line services. For this reason, we propose
a different model in which the users will generate and share anonymised
data-sets, and sell them on a distributed marketplace. We also propose one
specific application that exploits the availability of data to give to the users
feedback of various types on how the services they navigate respect their
privacy, or expose data to data trackers.

The architecture that we proposed is the first step towards the implementation
of our vision, and still, there are many unsolved research issues we need to
address before the system can be really implemented. In the paper, starting from
our background and the current state of the art, we propose some research
directions that must be followed to make IPPO reality.

Further work on the topic may address implementation issues and feasibility of IPPO,
by evaluating computational and storage costs of the proposed system, in order to
analyse the efficiency of the proposed system, compared to current solutions.

\bibliographystyle{elsarticle-num}
\bibliography{bibliography}

\begin{thebibliography}{10}
\bibitem{cadwalladr2017great}
C.~Cadwalladr, The great british brexit robbery: how our democracy was
  hijacked, The Guardian 20.
\bibitem{Peck2016Blockchain}
M.~Peck, A blockchain currency that beat s bitcoin on privacy, IEEE Spectrum
  53~(12).
\end{thebibliography}

\end{document}